\documentclass[
twocolumn,aip,jcp,showpacs,showkeys,preprintnumbers,amsmath,amssymb,reprint]
{revtex4-1}

\usepackage{graphicx}
\usepackage{dcolumn}
\usepackage{bm}
\usepackage{color}
\usepackage{threeparttable}
\usepackage{multirow}

\begin{document}

\title{%
NMR shieldings from density functional perturbation theory:\\
GIPAW versus all-electron calculations.
}

\author{G.~A.~de Wijs}
\affiliation{%
Radboud University, Institute for Molecules and Materials,
Heyendaalseweg 135, NL-6525 AJ Nijmegen, The Netherlands
}

\author{R.~Laskowski}
\affiliation{%
Institute of High Performance Computing, A$\ast$STAR, 1 Fusionopolis Way, \#16-16, Connexis, Singapore 138632
}

\author{P.~Blaha}
\affiliation{%
Institute of Materials Chemistry, Vienna University of Technology,
Getreidemarkt 9/165-TC, A-1060 Vienna, Austria
}

\author{R.~W.~A.~Havenith}
\affiliation{%
Zernike Institute for Advanced Materials, Stratingh Institute for Chemistry, University of Groningen,
Nijenborgh~4, NL-9747~AG Groningen, The Netherlands
}
\affiliation{%
Ghent Quantum Chemistry Group, Department of Inorganic and Physical Chemistry,
Ghent University, Krijgslaan 281 (S3), B-9000 Gent, Belgium
}

\author{G.~Kresse}
\author{M.~Marsman}
\affiliation{%
Faculty of Physics and Center for Computational Materials Science,
University of Vienna, Sensengasse 8/12, A-1090 Vienna, Austria
}

\date{\today}

\begin{abstract}
We present a benchmark of the density functional linear response calculation of
NMR shieldings within the Gauge-Including Projector-Augmented-Wave method against
all-electron Augmented-Plane-Wave$+$local-orbital and uncontracted Gaussian basis set results for NMR shieldings in
molecular and solid state systems.
In general, excellent agreement between the aforementioned methods is obtained.
Scalar relativistic effects are shown to be quite large for nuclei in molecules in the deshielded limit.
The small component makes up a substantial part of the relativistic corrections.
\end{abstract}


\maketitle

\section{\label{sec:introduction}Introduction}

Nuclear magnetic resonance (NMR) spectroscopy presents a powerful and sensitive
probe of the structure of molecules, liquids, and solids on the atomic scale. 
In general, however, the retrieval of structural information from measured NMR spectra
is a far from trivial process, since as yet, empirical rules that map between the
NMR spectrum and the structure were found to exist only for relatively simple
organic molecules.
To connect the features of measured NMR spectra unambiguously to complex structural
properties therefore remains difficult (and is often impossible) without additional
input from {\it ab initio} quantum mechanical modeling.

In the case of molecular systems and finite clusters of atoms, the {\it ab initio}
techniques traditionally used in quantum chemistry have been successfully
applied to aid in the analysis of experimental solution-state NMR spectra for
quite some time now.~\cite{helgaker:cr:1999}
In the case of solid-state NMR, finite clusters of atoms were used to approximate 
the infinite solid.
{\it Ab initio} quantum mechanical calculations of NMR shieldings in
truly extended systems under periodic boundary conditions were first performed
by Mauri, Pfrommer, and Louie,\cite{mauri:prl:1996} using a linear response approach.

Although Mauri {\it et al.}~derived their expressions starting from an
all-electron Hamiltonian, practical implementations thereof used norm-conserving
pseudopotentials, which largely limited its applicability to the calculation of
chemical shifts for light elements.
Only with the introduction of the Gauge-Including Projector Augmented Wave (GIPAW)
method by Pickard and Mauri,\cite{pickard:prb:2001} and its extension to non-norm-conserving
pseudo orbitals by Yates, Pickard, and Mauri (YPM)\cite{yates:prb:2007} several years later,
did the calculation of NMR shieldings become routinely possible for most of the nuclei
commonly studied in NMR.

The GIPAW method permits to obtain accurate chemical shielding with a
plane-wave basis set.
As in the original projector-augmented-wave method of Bl\"{o}chl,\cite{bloechl:prb:1994}
it recovers the shape of the all-electron Kohn-Sham orbitals near the nucleus 
through an augmentation procedure involving atom-centered functions.
In addition, the GIPAW method solves the gauge problem arising from incompleteness of the
atom-centered augmentation functions in a way similar as done for molecules in the
Gauge Independent Atomic Orbital (GIAO) method.~\cite{ditchfield:mp:1974}
The GIPAW formalism of YPM has been implemented in several plane-wave codes
(for instance, CASTEP,\cite{castep} Quantum Espresso,\cite{espresso} and PARATEC)
and is currently widely used in the solid-state NMR community for an extensive range
of applications (see, {\it e.g.}, Ref.~\onlinecite{charpentier:ssnmr:2011} and references therein).

Recently the calculation of NMR shieldings within the augmented-plane-wave~$+$~local-orbital
(APW$+$lo) method was implemented in WIEN2k.~\cite{laskowski:prb:2012,laskowski:prb:2013,laskowski:prb:2014}
In this paper we benchmark the recent implementation (by several of us) of the
linear response calculation of NMR shieldings within the GIPAW formalism of YPM
in the Vienna {\it Ab initio} Simulation Package (VASP)\cite{vasconcelos:jcp:2013}
against all-electron APW$+$lo results for NMR shieldings in molecular and solid state systems
and against non-relativistic LCAO calculations using DALTON (Ref.~\onlinecite{dalton}) and large uncontracted Gaussian
basis sets.
These benchmarks serve to further validate the aforementioned implementation
in VASP as well as WIEN2k, and as importantly, to establish the quality of the GIPAW approach
and the parameterization of the atomic scattering properties involved.

The rest of this paper is organized as follows:
in section~\ref{sec:theory} we reiterate the linear response expressions
for the NMR shieldings in the GIPAW formalism.
The particulars of the PAW data sets we use in our GIPAW NMR benchmarks and
the general setup of these calculations are discussed in Sec.~\ref{sec:setup}.
The results of aforementioned GIPAW benchmarks are presented in Sec.~\ref{sec:results}
and compared to all-electron calculations and---partly---to experiment.
Conclusions are drawn in Sec.~\ref{sec:conclusions}.

\section{\label{sec:theory}Theory}

The magnetic shielding tensor $\sigma({\bf R})$ at nuclear position ${\bf R}$ is
found from the ratio of the induced magnetic field at the aforementioned position
to an externally applied magnetic field ${\bf B}$:
\begin{equation}
\sigma_{\alpha\beta}({\bf R})=-\frac{\partial B^\mathrm{ind}_{\alpha}({\bf R})}{\partial B_{\beta}} .
\label{eq:shielding}
\end{equation}
The induced magnetic field ${\bf B}_\mathrm{ind}$ is given by the Biot-Savart law,
\begin{equation}
{\bf B}_\mathrm{ind}({\bf R})=\frac{1}{c}\int {\bf j}_\mathrm{ind}({\bf r})\mbox{\boldmath $\times$}\frac{{\bf R}-{\bf r}}{|{\bf R}-{\bf r}|^3} d{\bf r},
\label{eq:bind}
\end{equation}
where ${\bf j}_\mathrm{ind}$ is the current induced by the external magnetic field ${\bf B}$.
The induced current is commonly calculated from the linear response of the system to the external magnetic field.

To first-order in the external magnetic field, ${\bf j}_\mathrm{ind}$ is given by
\begin{align}
{\bf j}_\mathrm{ind}({\bf r})=\sum_{i \in \mathrm{occ}} & \Bigl(
\langle\psi^{(1)}_i|{\bf J}^p({\bf r})|\psi^{(0)}_i\rangle +
 \langle\psi^{(0)}_i|{\bf J}^p({\bf r})|\psi^{(1)}_i\rangle \nonumber\\
&+\langle\psi^{(0)}_i|{\bf J}^d({\bf r})|\psi^{(0)}_i\rangle
\Bigr),
\label{eq:jind}
\end{align}
where 
\begin{equation}
{\bf J}^p({\bf r})=-\frac{{\bf p}|{\bf r}\rangle\langle{\bf r}|+|{\bf r}\rangle\langle{\bf r}|{\bf p}}{2}
\label{eq:jpara}
\end{equation}
and
\begin{equation}
{\bf J}^d({\bf r})=-\frac{{\bf B}\mbox{\boldmath $\times$}{\bf r}}{2c}|{\bf r}\rangle\langle{\bf r}|
\label{eq:jdia}
\end{equation}
are the paramagnetic and diamagnetic current operators, respectively.
In Eq.~(\ref{eq:jind}), $\psi^{(0)}$ denote ground state orbitals, {\it i.e.}, the solutions to
\begin{equation}
H^{(0)}|\psi^{(0)}_i\rangle = \epsilon^{(0)}_i |\psi^{(0)}_i\rangle,
\label{eq:ks}
\end{equation}
and $\psi^{(1)}$ the first-order
response of the orbitals to the external magnetic field.
The sum in Eq.~(\ref{eq:jind}) goes over all occupied states.

In the symmetric gauge, the perturbation of the Hamiltonian to first-order in the external magnetic field,
is given by
\begin{equation}
H^{(1)}=\frac{1}{2c}{\bf L}\cdot{\bf B}.
\end{equation}
With the above definition the first-order change in the orbitals is straightforwardly found to be
\begin{equation}
|\psi^{(1)}_i\rangle= \mathcal{G}(\epsilon^{(0)}_i) H^{(1)}|\psi^{(0)}_i\rangle,
\label{eq:green}
\end{equation}
where $\mathcal{G}$ is the Green's function
\begin{equation}
\mathcal{G}(\epsilon)=\sum_{j \in \mathrm{vir}}
\frac{|\psi^{(0)}_j\rangle\langle\psi^{(0)}_j|}{\epsilon-\epsilon^{(0)}_j}
\end{equation}
and the sum is over all empty (virtual) orbitals.
Commonly, the sum over empty orbitals is avoided by recasting Eq.~(\ref{eq:green}) as a Sternheimer equation:
\begin{equation}
\left(\epsilon^{(0)}_i-H^{(0)}\right)|\psi^{(1)}_i\rangle=P_c H^{(1)}|\psi^{(0)}_i\rangle,
\label{eq:sternheimer}
\end{equation}
to be solved for $\psi^{(1)}_i$.
In the above $P_c = 1 - \sum_{i \in \mathrm{occ}} |\psi^{(0)}_i\rangle\langle\psi^{(0)}_i|$ represents a projection onto the virtual subspace.

In plane wave based implementations, Eq.~(\ref{eq:bind}) is most conveniently evaluated
in reciprocal space,
\begin{equation}
{\bf B}_\mathrm{ind}({\bf R})=\frac{4\pi i}{c}\sum_{{\bf G}\neq 0}
\frac{{\bf G}\mbox{\boldmath $\times$} {\bf j}_\mathrm{ind}({\bf G})}{G^2}e^{i{\bf G}\cdot{\bf R}} ,
\label{eq:g}
\end{equation}
where $\bf G$ are the reciprocal space vectors.
In solid state systems there is an additional contribution at ${\bf G}=0$, {\it i.e.},
a uniform field, that is determined by the shape of the sample and the macroscopic
magnetic susceptibility tensor $\chi$.
For a spherical sample this contribution is given by:
\begin{equation}
{\bf B}_\mathrm{ind}({\bf G}=0)=\frac{8\pi}{3}\overset\leftrightarrow\chi{\bf B}.
\label{eq.bg0}
\end{equation}

The magnetic susceptibility tensor may be numerically calculated as proposed
by Mauri and Louie,\cite{louie:prl:1996}
\begin{equation}
\overset\leftrightarrow\chi=\lim_{q \rightarrow 0}
\frac{\overset\leftrightarrow F(q)-2\overset\leftrightarrow F(0)+\overset\leftrightarrow F(-q)}{q^2},
\label{eq:chi}
\end{equation}
where $F_{\alpha\beta}(q)=(2-\delta_{\alpha\beta})Q_{\alpha\beta}(q)$,
$\alpha,\beta=x,y,z$ are the cartesian directions, and the tensor $Q(q)$ can
be written as
\begin{multline}
\overset\leftrightarrow Q (q)= \\ \frac{1}{N_{\bf k}\Omega c^2}\sum_{n{\bf k} \in \mathrm{occ}}\sum^{x,y,z}_{\gamma}
\mathrm{Re}\{\langle u^{(0)}_{n{\bf k}}|{\bf A}_\gamma \mathcal{G}_{{\bf k}+q\hat{\gamma}}(\epsilon^{(0)}_{n{\bf k}})
{\bf A}_\gamma|u^{(0)}_{n{\bf k}}\rangle\}.
\label{eq:qq}
\end{multline}
with
\begin{equation}
{\bf A}_\gamma=\hat{\bf u}_\gamma \mbox{\boldmath $\times$} ({\bf p}+{\bf k}).
\label{eq:a}
\end{equation}
In Eq.~(\ref{eq:qq}), the functions $u^{(0)}_{n{\bf k}}$ denote the cell periodic part of the ground state
Bloch orbitals, $N_{\bf k}$ is the number of $k$-points chosen to sample the first Brillouin zone,
$\Omega$ the volume of the unit cell, and the sum over $n$ and ${\bf k}$ includes all occupied Bloch orbitals.
The Green's function in Eq.~(\ref{eq:qq}) is given by
\begin{equation}
\mathcal{G}_{\bf k}(\epsilon)=\sum_{n \in \mathrm{vir}}
\frac{|u^{(0)}_{n{\bf k}}\rangle\langle u^{(0)}_{n{\bf k}}|}{\epsilon-\epsilon^{(0)}_{n{\bf k}}}.
\label{eq:gnk}
\end{equation}

In practice, we have implemented Eqs.~(\ref{eq:shielding})-(\ref{eq:gnk})
within the Gauge-Including Projector-Augmented-Wave (GIPAW) method of Yates, Pickard,
and Mauri (YPM).\cite{yates:prb:2007}
The GIPAW deals with several numerical issues that plague the PAW method (in a uniform magnetic field):
(a) it reestablishes the translational symmetry that is broken by the PAW method (see next section),\cite{mauri:prl:1996}
(b) it balances the different rates of convergence of the para- and diamagnetic contributions to the induced current 
-- which also affects translational symmetry -- {\it via} the generalized $f$-sum rule,\cite{mauri:prl:1996}
and (c) it solves the position operator problem with the help of a reciprocal space modulation vector ${\bf q}$ (cf.~Eq.~\ref{eq:chi}).

\subsection{\label{sec:theory:gipaw}GIPAW}

In the Projector-Augmented-Wave (PAW) method of Bl\"{o}chl the one-electron
orbitals $\psi_n$ are derived from pseudo (PS) orbitals $\widetilde{\psi}_n$
by means of a linear transformation~\cite{bloechl:prb:1994}
\begin{equation}
|\psi_n\rangle = \mathcal{T}|\widetilde{\psi}_n\rangle
\label{eq:paw1}
\end{equation}
with 
\begin{equation}
\mathcal{T}=1+\sum_j 
\left( |\phi_j\rangle - |\tilde{\phi}_j\rangle\right)\langle\tilde{p}_j|.
\label{eq:paw2}
\end{equation}
The PS orbitals $\widetilde{\psi}_n$ are the variational quantities of
the PAW method and are expanded in plane waves.
The additional local basis functions, $\phi_j$ and
$\widetilde{\phi}_j$, are non-zero only within non-overlapping spheres
centered at the atomic sites ${\bf R}_j$, the so-called PAW spheres.
In the interstitial region between the PAW spheres, therefore, the true
one-electron orbitals ${\psi}_n$ are identical to the PS orbitals
$\widetilde{\psi}_n$.
Inside the spheres the PS orbitals are only a computational tool and a
bad approximation to the true orbitals, since not even the
norm of the true orbital is reproduced.
In all practical implementations of the PAW method, the all-electron
(AE) partial-waves $\phi_j$ are chosen to be solutions of the
spherical (scalar relativistic) Schr\"{o}dinger equation for a
\textit{non-spinpolarized atom} at a specific energy $\varepsilon_j$, and for a specific angular momentum $l_j$.
The pseudo partial waves $\widetilde{\phi}_j$ are equivalent to their
AE counterparts outside a core radius $r_c$ and match continuously onto
$\phi_j$ inside this radius. In the PAW data sets distributed with VASP they
are constructed in accordance with a revised Rappe,
Rabe, Kaxiras, and Joannopoulos (RRKJ) scheme.~\cite{rappe:prb:1990,kresse:jpcm:1994}
The projector functions $\widetilde{p}_j$ are constructed to be dual
to the PS partial waves, i.e.,
\begin{equation}
\label{eq:proj}
\langle\widetilde{p}_j|\widetilde{\phi}_{j'}\rangle=
\delta_{jj'}.
\end{equation}
A detailed construction recipe for the projector functions can be
found in Ref.~\onlinecite{kresse:jpcm:1994}.
For a comprehensive introduction to the PAW method we refer the reader
to the seminal paper of Bl\"{o}chl (Ref.~\onlinecite{bloechl:prb:1994}) and the work of
Kresse and Joubert.~\cite{kresse:prb:1999}
 
In a uniform magnetic field {\bf B} there is an additional complication as the ground state orbitals acquire an
additional phase factor upon translation over a vector {\bf t}, in accordance with:
\begin{equation}
\langle {\bf r}|\psi_n^{\bf t}\rangle=
e^{\frac{i}{2c}{\bf r}\cdot{\bf t}\mbox{\boldmath $\times$}{\bf B}}
\langle {\bf r}-{\bf t} | \psi_n \rangle
\end{equation}
(in the symmetric gauge).
This additional phase factor causes a very slow convergence of the
linear transformation of Eq.~\ref{eq:paw1} with respect to the number
of projectors $\tilde{p}_j$.
To solve this problem, Pickard and Mauri introduced the
so-called Gauge-Including PAW transformation that includes the aforementioned
phase factor explicitly:~\cite{pickard:prb:2001,yates:prb:2007}
\begin{equation}
\bar{\mathcal{T}}=1+\sum_j 
e^{\frac{i}{2c}{\bf r}\cdot{\bf R}_j\mbox{\boldmath $\times$}{\bf B}}
\left( |\phi_j\rangle - |\tilde{\phi}_j\rangle\right)\langle\tilde{p}_j|
e^{-\frac{i}{2c}{\bf r}\cdot{\bf R}_j\mbox{\boldmath $\times$}{\bf B}}.
\label{eq:t_gipaw}
\end{equation}
Using the transformation of Eq.~\ref{eq:t_gipaw}, it is straightforward to
show that with any local operator $O$ acting on $\psi_n$ the GIPAW associates
a PS operator $\bar{O}$ acting on the PS orbitals $\bar{\psi}_n$:
\begin{equation}
\bar{O}=O+\sum_{jj'} |\bar{p}_j\rangle \left( 
\langle\bar{\phi}_j|O|\bar{\phi}_{j'}\rangle -
\langle\bar{\tilde{\phi}}_j|O|\bar{\tilde{\phi}}_{j'}\rangle
\right)\langle\bar{p}_{j'}|
\label{GIPAWoperator}
\end{equation}
where
\begin{eqnarray}
|\bar{p}_j\rangle &=&
e^{\frac{i}{2c}{\bf r}\cdot{\bf R}_j\mbox{\boldmath $\times$}{\bf B}}|\tilde{p}_j\rangle, \nonumber \\
|\bar{\phi}_j\rangle &=&
e^{\frac{i}{2c}{\bf r}\cdot{\bf R}_j\mbox{\boldmath $\times$}{\bf B}}|\phi_j\rangle, \\
|\bar{\tilde{\phi}}_j\rangle &=&
e^{\frac{i}{2c}{\bf r}\cdot{\bf R}_j\mbox{\boldmath $\times$}{\bf B}}|\tilde{\phi}_j\rangle. \nonumber
\end{eqnarray}

As shown by YPM, to first-order in the magnetic field ${\bf B}$,
the GIPAW transformation of the induced current is given by
\begin{align}
{\bf j}^{(1)}_\mathrm{ind}({\bf r})=
\sum^\mathrm{occ}_i&\Bigl(
2\mathrm{Re}\{\langle\bar{\psi}^{(0)}_i|\bar{\bf J}^{(0)}({\bf r})|\bar{\psi}^{(1)}_i\rangle\} \nonumber \\
&-\sum^\mathrm{occ}_j\langle\bar{\psi}^{(0)}_i|\bar{\bf J}^{(0)}({\bf r})|\bar{\psi}^{(0)}_j\rangle
\langle\bar{\psi}^{(0)}_j|\bar{S}^{(1)}|\bar{\psi}^{(0)}_i\rangle \nonumber \\
&+\langle\bar{\psi}^{(0)}_i|\bar{\bf J}^{(1)}({\bf r})|\bar{\psi}^{(0)}_i\rangle 
\Bigr)
\label{eq:jind_gipaw}
\end{align}
In the above, $\bar{\psi}^{(0)}$ are the ground state orbitals, {\it i.e.}, the solutions to
\begin{equation}
\bar{H}^{(0)}|\bar{\psi}^{(0)}_n\rangle =
\epsilon^{(0)}_n\bar{S}^{(0)}|\bar{\psi}^{(0)}_n\rangle.
\end{equation}
This equation is
the GIPAW transform of the Kohn-Sham equations (see Eq.~\ref{eq:ks}) to zeroth-order
in the magnetic field, so it is just the usual PAW generalized Kohn-Sham eigenvalue equation
({\it i.e.}, $\bar{H}^{(0)}$ and $\bar{S}^{(0)}$ are equal to the PAW Hamiltonian
and overlap operators of Refs.~\onlinecite{bloechl:prb:1994}
and~\onlinecite{kresse:prb:1999}, and consequently $\bar{\psi}^{(0)}_n=\tilde{\psi}_n$
and $\epsilon^{(0)}_n=\epsilon_n$).

The current operators, to zeroth- and first-order in the magnetic field,
are given by
\begin{equation}
\bar{{\bf J}}^{(0)}={\bf J}^p({\bf r})+\sum_{\bf R} \Delta {\bf J}^p_{\bf R}({\bf r})
\label{eq:jpara_gipaw}
\end{equation}
and
\begin{equation}
\bar{{\bf J}}^{(1)}={\bf J}^d({\bf r})+\sum_{\bf R} \left(\Delta {\bf J}^d_{\bf R}({\bf r})
+\frac{1}{2ci}\left[{\bf B}\mbox{\boldmath $\times$}{\bf R}\cdot{\bf r},\Delta {\bf J}^p_{\bf R}({\bf r})\right]\right),
\label{eq:jdia_gipaw}
\end{equation}
respectively.
These are easily recognized as the paramagnetic and diamagnetic current operators of
Eqs.~\ref{eq:jpara} and~\ref{eq:jdia}, plus additional one-center correction terms (the terms involving
$\Delta {\bf J}^p_{\bf R}$ and $\Delta {\bf J}^d_{\bf R}$; see Ref.~\onlinecite{yates:prb:2007}).

For non-normconserving PAW data sets the induced current of Eq.~(\ref{eq:jind_gipaw}) contains an additional
contribution connected to the first-order change of the GIPAW orbital overlap operator with respect to
the magnetic field,
\begin{equation}
\bar{S}^{(1)}=\frac{1}{2c}\sum_{\bf R} {\bf R}\mbox{\boldmath $\times$}\frac{1}{i}\left[{\bf r},Q_{\bf R}\right]\cdot{\bf B}.
\end{equation}

The first-order change in the GIPAW wave functions $\bar{\psi}^{(1)}$,
is found by solving a generalized Sternheimer equation,
\begin{equation}
\left(\epsilon^{(0)}_i \bar{S}^{(0)}-\bar{H}^{(0)}\right)|\bar{\psi}^{(1)}_i\rangle=
P_c \left(\bar{H}^{(1)}-\epsilon^{(0)}_i\bar{S}^{(1)}\right)|\bar{\psi}^{(0)}_i\rangle,
\label{eq:sternheimer_gipaw}
\end{equation}
where
\begin{equation}
P_c=1-\sum_{i \in \mathrm{occ}} \bar{S}^{(0)}|\bar{\psi}^{(0)}_i\rangle\langle\bar{\psi}^{(0)}_i|,
\end{equation}
and
\begin{equation}
\bar{H}^{(1)}=\frac{1}{2c}\left(
{\bf L} + \sum_{\bf R} {\bf R}\mbox{\boldmath $\times$}\frac{1}{i}\left[{\bf r},V^\mathrm{nl}_{\bf R}\right]+
\sum_{\bf R} {\bf L}_{\bf R}Q_{\bf R}
\right)\cdot{\bf B},
\end{equation}
is the first-order contribution to the GIPAW Hamiltonian.

The macroscopic magnetic susceptibility is calculated in accordance with the {\it ansatz}
of YPM [see Eqs.~(47) and~(48) of Ref.~\onlinecite{yates:prb:2007}], which equals
Eqs.~(\ref{eq:chi})-(\ref{eq:gnk}) for $\mathcal{T}=1$, but represents an approximation
otherwise.
We will not repeat the expressions here.

\subsection{\label{sec:theory:core}Core contributions}

As was shown by Gregor, Mauri, and Car~\cite{gregor:jcp:1999} the contribution
of the core electrons to the NMR shieldings is essentially rigid,
and can be calculated from the atomic orbitals of the core electrons:
\begin{equation}
\sigma^c_{\alpha\beta}({\bf R})=\frac{1}{2c}
\sum^\mathrm{core}_i \langle \psi_i | \frac{1}{r} | \psi_i \rangle \delta_{\alpha\beta},
\label{eq:sigmac}
\end{equation}
where the sum is understood to be taken over the core electronic states at atomic site
${\bf R}$,  and the delta function expresses the fact that the core electrons only contribute
isotropically. Of course, in the PAW formalism, we use frozen core states.

As is usual in the
GIPAW, the excitations from the valence to the core states are not included
in the Green's function (the pseudo equivalent of Eq.~\ref{eq:gnk}). In principle these
should be included in the proper decoupling of the valence and core contributions to the
chemical shieldings, however, in IGAIM (Individual Gauges for Atoms In Molecules) and similar
methods their neglect gives rise to errors much smaller than one ppm.\cite{gregor:jcp:1999}
Moreover, such inaccuracies, should they play a role, can be minimized by unfreezing the
shallowest core shell(s).

The contribution of the core electrons to the macroscopic magnetic susceptibility
is only approximately rigid.~\cite{louie:prl:1996}
It is commonly assumed to be rigid, though, and included as:
\begin{equation}
\chi^c_{\alpha\beta}=-\frac{1}{\Omega c^2}
\sum^\mathrm{core}_i \langle \psi_i | r^2 | \psi_i \rangle \delta_{\alpha\beta},
\label{eq:chic}
\end{equation}
where the sum is now taken over all core electronic states of the system.

\section{\label{sec:setup}Computational setup}

All calculations employed the Perdew, Burke and Ernzerhof generalized gradient
expansion.\cite{pbe1,pbe2}

\subsection{GIPAW}
\label{subectGIPAW}

We carried out two series of GIPAW calculations:
one with the standard data sets that are designed for general use,\cite{vasp:manual}
and one where we selected data sets to aim for high accuracy of the shieldings.
The former, referred to as ``standard'' or ``stand'' below,
are a compromise between many demands and should
yield good performance for a reasonable plane wave cutoff energy.
They are not expected to be optimal for calculating shieldings.
Accurate shieldings require an accurate PAW reconstruction of states
in the immediate vicinity of the nucleus, as currents in this region have a high
impact on the field at the nucleus.
Such accuracy is only needed in the calculation of few properties,
that are typically related to spectroscopic techniques probing
the nucleus, {\it e.g.} for electric field gradients and NMR.
The latter series, referred to as ``optimal'',
is intended to give high accuracy for the shieldings.
Details of the data sets are compiled in Table~\ref{tab:pawdata}.
Completeness in the projectors and partial waves is easier to
realize when the pseudo partial waves are norm-conserving and
have an extra (radial) node for each additional projector
(within the same angular momentum channel).
Logically this results in harder projector functions.
This is the case for the $\ast$\_sv\_GW\_nc data sets (see Ref.~\onlinecite{klimes:prb:2014}).
Typically these also have a substantial part of the core states unfrozen,
which helps in keeping the PAW matching radii small.
Alternatively, for some elements we used just very hard data sets ($\ast$\_h).
For the ``standard'' and ``optimal'' series we used kinetic energy cutoffs
of 700 and 900~eV respectively.

Inside the muffin tin spheres, WIEN2k uses a basis that consists of solutions to the
scalar-relativistic Schr\"odinger equation for the spherical atom: these atomic orbitals have a
so-called ``large'' ($A$) and ``small'' ($B$) component.
The AE-partial waves of the VASP PAW data sets are solutions to the same
scalar-relativistic equation.
However, in the VASP PAW data sets the ``small''-component $B$ is not retained.
Instead the large component $A$ is rescaled to have the correct norm
($\sqrt{\langle A| A\rangle+\langle B| B\rangle}$).
Since the $B$-component becomes appreciable close to the nuclei,
such a treatment causes non-negligible errors in the NMR shielding.
This turns out to be problematic for some of the molecular systems. 

To solve this issue, we recompute the $B$-component on-the-fly,
but use it only to evaluate the AE one-centre contributions to the shielding:
the $B$ AE-partial wave is reconstructed as the radial derivative
of the $A$ AE-partial wave, rescaled with a ZORA-like expression
for the relativistic mass.\cite{zora}
Equation.~\ref{GIPAWoperator} is modified and becomes:
\begin{multline}
\bar{O}=O+ \\ \sum_{jj'} |\bar{p}_j\rangle \left( 
\langle\bar{\phi}_j^A|O|\bar{\phi}_{j'}^A\rangle +
\langle\bar{\phi}_j^B|O|\bar{\phi}_{j'}^B\rangle -
\langle\bar{\tilde{\phi}}_j|O|\bar{\tilde{\phi}}_{j'}\rangle
\right)\langle\bar{p}_{j'}|
\end{multline}
The $A$ and $B$ AE-partial waves are renormalized such that the
electron count in each channel inside the sphere is unaffected.

Note that the aforementioned issue only applies to the contributions to the chemical
shielding stemming from the {\it valence} electronic states.
The core contributions to the chemical shieldings always include the contributions
of the ``small'' component explicitly in both, VASP as well as WIEN2k.

\subsection{APW}

The WIEN2k calculations performed in this work apply the formalism
described in
Refs.~\onlinecite{laskowski:prb:2012,laskowski:prb:2014}. The standard
APW basis set is extended with eight additional local orbitals (NMR-LOs)
at higher expansion energies for all ``chemical'' $l + 1$ angular momenta
using a procedure described in Ref.~\onlinecite{laskowski:prb:2012}. The
Greens function used to represent the perturbation of the ground state
is augmented with additional $r\frac{\partial}{\partial r} u$ terms
in order to accelerate convergence with respect to the number of
NMRLOs.~\cite{laskowski:prb:2014}  The separation of the valence and
core states substantially affects the absolute values of the
shielding,\cite{laskowski:prb:2014} thus in the current work we apply
the corresponding core-correction.  All molecular calculations are
done with only the 1s state as core for all atoms (except H) because
of the short bond-lengths in these molecules, while for the bulk
calculations we applied the usual WIEN2k criterion defining valence states
as states with atomic eigenvalues above  $-6$~Ry.
Core-states are treated fully-relativistically, but in the
self-consistent spherical potential only and the corresponding
NMR-shielding is calculated {\it via} Eq.~\ref{eq:sigmac}. The numerical
parameters are set to standard WIEN2k values. The convergence with
respect to the basis set size (RKMAX) has been tested and the presented
values of the shielding for the molecular systems (in 
Table~\ref{tab:molecular_shifts_dalton})
are extrapolated to infinite RKMAX (typically extrapolation shifts the 
shielding by 1-3 ppm compared to the largest applied RKMAX).

\subsection{GIAO}

For molecules calculations were carried out with the quantum chemical
code DALTON (Refs.~\onlinecite{dalton,daltondft}) using uncontracted aug-cc-pCVXX (XX=DZ, TZ, QZ, 5Z) basis
sets (unless stated otherwise).\cite{dunning1,dunning2,dunning3,dunning4,dunning5,dunning6,dunning7,dunning8,emsl1,emsl2}
Uncontracting the basis sets is crucial to observe a convergence of the shieldings (and a lowering of
the total energy) with increasing basis set quality. These calculations were non-relativistic.

\begin{table*}
\begin{threeparttable}
\caption{\label{tab:pawdata}Frozen core configurations and matching core radii $r_c$ for $s$, $p$, $d$ and $d$ partial waves in atomic units for the optimal PAW data sets.
The number of projectors for each quantum number $\ell$ is given in parenthesis. An ``$l$'' indicates this is taken as local potential.}
\begin{tabular}{lldddd}
\hline
\hline
               & frozen core       & \multicolumn{1}{c}{$s$}     & \multicolumn{1}{c}{$p$}     & \multicolumn{1}{c}{$d$}      & \multicolumn{1}{c}{$f$}     \\
\hline
H              &                   & 1.10~(2) & 1.10~(1) &     ~     &     ~    \\
Li\_sv         &                   & 1.40~(1) ~~ 1.70~(1) & 1.40~(1) & 1.40~(1)l &     ~    \\
%              &                   & 1.70~(1) &          &           &     ~    \\
Be\_sv         &                   & 1.50~(3) & 1.80~(2) & 1.80~(1)l &     ~    \\
C\_h           & [He]              & 1.10~(2) & 1.10~(2) & 1.10~(1)l &     ~    \\
N\_h           & [He]              & 1.10~(2) & 1.10~(2) & 1.10~(1)l &     ~    \\
O\_h           & [He]              & 1.10~(2) & 1.10~(2) & 1.10~(1)l &     ~    \\
F\_h           & [He]              & 0.85~(2) & 1.10~(2) & 1.10~(1)l &     ~    \\
Na\_sv\_GW\_nc & [He]              & 1.20~(2) & 2.20~(3) & 2.20~(2)  &     ~    \\
Mg\_sv\_GW\_nc & [He]              & 1.15~(1) ~~ 1.30~(2) & 1.65~(3) & 1.65~(2)  &     ~    \\
%              &                   & 1.30~(2) &          &           &     ~    \\
Al\_sv\_GW\_nc & [He]              & 1.75~(3) & 2.00~(3) & 1.80~(2)  & 2.00~(1) \\
Si\_sv\_GW\_nc & [He]              & 1.70~(3) & 1.95~(3) & 1.70~(2)  & 2.00~(1) \\
Si\_sv\_GW\_nc & [He]              & 1.70~(3) & 1.95~(3) & 1.70~(2)  & 2.00~(1) \\
P\_sv\_GW\_nc  & [He]              & 1.70~(3) & 1.95~(3) & 1.70~(2)  & 2.00~(1) \\
Cl\_GW\_nc     & [Ne]              & 1.14~(2) & 1.25~(2) & 1.70~(2)  & 1.70~(1)l \\
K\_sv\_GW\_nc  & [Ne]              & 0.95~(2) & 1.76~(3) & 2.10~(2)  & 2.10~(2) \\
Ca\_sv\_GW\_nc & [Ne]              & 0.90~(2) & 1.65~(3) & 1.90~(3)  & 2.10~(1) \\
Ti\_sv\_GW\_nc & [Ne]              & 0.85~(2) & 1.41~(2) & 1.90~(3)  & 1.90~(2) \\
Ga\_sv\_GW\_nc & [Ne]              & 1.23~(2) ~~ 1.55~(1) & 1.70~(3) & 1.90~(2)  & 1.90~(2) \\
%              &                   & 1.55~(1) &          &           &     ~    \\
Rb\_sv\_GW\_nc & [Ar]$(3d)^{10}$   & 1.16~(2) & 2.10~(3) & 2.30~(3)  & 2.10~(2) \\
Sr\_sv\_GW\_nc & [Ar]$(3d)^{10}$   & 1.10~(2) & 2.00~(3) & 2.30~(3)  & 2.10~(2) \\
Zr\_sv\_GW\_nc & [Ar]$(3d)^{10}$   & 1.01~(2) & 1.90~(3) & 2.10~(3)  & 1.90~(2) \\
In\_sv\_GW\_nc & [Ar]$(3d)^{10}$   & 1.66~(2) ~~ 1.80~(1) & 2.00~(3) & 2.20~(3)  & 1.90~(2) \\
%              &                   & 1.80~(1) &          &           &     ~    \\
Sn\_sv\_GW\_nc & [Ar]$(3d)^{10}$   & 1.60~(2) ~~ 1.70~(1) & 2.00~(3) & 2.20~(3)  & 1.90~(2) \\
%              &                   & 1.70~(1) &          &           &     ~    \\
Cs\_sv\_GW\_nc &
[Ar]$(3d)^{10}(4p)^6(4d)^{10}$     & 1.30~(1) ~~ 1.40~(2) & 2.25~(3) & 2.60~(3)  & 2.10~(2) \\
%              &                   & 1.40~(2) &          &           &     ~    \\
Ba\_sv\_GW\_nc &
[Ar]$(3d)^{10}(4p)^6(4d)^{10}$     & 1.30~(1) ~~ 1.40~(2) & 2.20~(3) & 2.50~(3)  & 2.10~(2) \\
%              &                   & 1.40~(2) &          &           &     ~    \\
Tl\_sv\_GW     & [Kr]$(4d)^{10}$   & 1.75~(3) & 1.90~(3) & 2.15~(3)  & 2.30~(2) \\
\hline
\hline
\end{tabular}
\end{threeparttable}
\end{table*}

\section{\label{sec:results}Results}

\subsection{\label{sec:results:molecules}Molecules}

\begin{table*}
\caption{\label{tab:molecular_shifts_dalton}Calculated absolute isotropic
chemical shieldings (in ppm).
VASP calculations in were carried out in large boxes (see text).
VASP ``optim'' results were obtained with Al\_sv\_nc, Si\_sv\_nc, P\_sv\_nc and
F\_h PAW data sets.
``NR'' denote non-relativistic calculations.
``$^+B$'' denotes scalar relativistic calculations with two-component KS-orbitals in the atomic/PAW spheres.
The DALTON calculations are non-relativistic using GIAOs and have uncontracted
basis sets.
Identical molecular geometries were used in the WIEN2k, VASP and DALTON calculations.
Four slightly differently deformed tetrahedral geometries were used for [AlH$_4$]$^-$.
``MAD'' denotes ``mean absolute deviation'' for the set of molecules that have WIEN2k results.
}
\begin{threeparttable}
\begin{tabular}{lddcdddcddddd}
\hline
\hline
 & \multicolumn{6}{c}{All-electron methods} & & \multicolumn{5}{c}{GIPAW} \\
\cline{2-7} \cline{9-13} 
 & \multicolumn{2}{c}{\multirow{1}{*}{WIEN2k}} & & \multicolumn{3}{c}{\multirow{1}{*}{DALTON2011, aug-cc-pCVXX}} & & \multicolumn{4}{c}{\multirow{1}{*}{VASP}} & \multicolumn{1}{c}{other} \\
\cline{2-3} \cline{5-7} \cline{9-12}
 & & & & \multicolumn{1}{c}{5Z} & \multicolumn{1}{c}{QZ} & \multicolumn{1}{c}{TZ} & & \multicolumn{1}{c}{optim.} & \multicolumn{1}{c}{optim.} & \multicolumn{1}{c}{optim.} & \multicolumn{1}{c}{stand.} & \multicolumn{1}{c}{} \\ %
 & \multicolumn{1}{c}{NR} & \multicolumn{1}{c}{$^+B$} & & \multicolumn{1}{c}{NR} & \multicolumn{1}{c}{NR} & \multicolumn{1}{c}{NR} & & \multicolumn{1}{c}{NR} & \multicolumn{1}{c}{$^+B$} &  &  &  \\ %
\hline
\multicolumn{6}{l}{Al shieldings\tnote{a}} & & & & & & & \\[1.2ex]             
\multirow{4}{*}{\mbox{[AlH$_4$]$^-$}}                                                                 &        &       & &  477.94 &  478.23 &  480.89 & &  477.50 &  476.08 &  477.89 &  483.93 &         \\        %
                                                                                                      &        &       & &  478.15 &  478.43 &  481.10 & &  477.71 &  476.30 &  478.11 &  484.21 &         \\        %
                                                                                                      &        &       & &  478.25 &  478.53 &  481.20 & &  477.84 &  476.42 &  478.23 &  484.31 &         \\        %
                                                                                                      &        &       & &  478.92 &  479.21 &  481.88 & &  478.47 &  477.07 &  478.87 &  484.97 &         \\        %
Al$_2$H$_6$                                                                                           &  402.8 & 401.2 & &  404.25 &  404.49 &  406.96 & &  404.60 &  402.30 &  404.56 &  411.13 &         \\        %
AlH$_3$                                                                                               &  249.7 & 245.9 & &  250.24 &  250.46 &  254.22 & &  251.03 &  246.84 &  250.06 &  258.54 &         \\[1.2ex] %
\multicolumn{6}{l}{Si shieldings\tnote{a}} & & & & & & & \\[1.2ex]                                                                                                                                                   
Si$_2$H$_6$                                                                                           &  436.9 & 433.9 & &  439.38 &  439.65 &  443.27 & &  439.98 &  436.29 &  439.21 &  443.01 &         \\        %
SiH$_4$                                                                                               &  431.6 & 429.5 & &  433.63 &  433.92 &  437.76 & &  434.30 &  430.94 &  433.88 &  437.68 &         \\        %
Si$_2$H$_4$                                                                                           &  235.9 & 230.2 & &  238.28 &  238.53 &  241.53 & &  239.55 &  233.01 &  237.38 &  240.56 &         \\        %
SiH$_2$                                                                                               & -533.4 &-545.4 & & -530.20 & -530.34 & -527.12 & & -526.34 & -538.36 & -528.67 & -523.91 &         \\[1.2ex] %
\multicolumn{6}{l}{P shieldings\tnote{a}} & & & & & & & \multicolumn{1}{c}{QE\tnote{b}} \\[1.2ex]                                                                                                                   
P$_4$                                                                                                 &  858.8 & 860.0 & &  861.56 &  861.68 &  862.31 & &  862.54 &  862.54 &  863.27 &  864.49 &  861.16 \\        %
PH$_3$                                                                                                &        &       & &  576.50 &  576.82 &  580.02 & &  575.77 &  572.80 &  575.73 &  579.61 &  577.47 \\        %
P$_2$H$_4$                                                                                            &  517.0 & 513.0 & &  520.11 &  520.42 &  523.61 & &  519.46 &  514.79 &  518.16 &  522.93 &  519.68 \\        %
H$_3$PO$_4$                                                                                           &  281.8 & 275.1 & &  284.65 &  285.00 &  289.31 & &  283.87 &  276.30 &  281.21 &  292.51 &  285.73 \\        %
PF$_3$                                                                                                &  153.9 & 145.7 & &  155.72 &  155.94 &  159.74 & &  156.05 &  146.87 &  152.94 &  156.58 &  158.02 \\        %
P$_2$                                                                                                 & -306.0 &-328.0 & & -301.79 & -301.32 & -298.10 & & -301.98 & -323.83 & -313.77 & -318.00 & -318.62 \\[1.2ex] %
\multicolumn{6}{l}{F shieldings\tnote{c}} & & & & & & & \multicolumn{1}{c}{CASTEP\tnote{c}} \\[1.2ex]                                                                                                       
CH$_3$F                                                                                               &        &       & &  451.65 &  451.74 &  452.26 & &  450.91 &  451.03 &  451.10 &  452.45 &  452.1  \\        %
HF                                                                                                    &  399.0 & 398.7 & &  399.98 &  400.21 &  401.02 & &  398.64 &  398.47 &  398.70 &  398.97 &  398.8  \\        %
C$_6$F$_6$                                                                                            &        &       & &  316.73 &  317.14 &  319.20 & &  314.99 &  314.15 &  314.65 &  313.72 &  310.6  \\        %
CH$_2$F$_2$                                                                                           &        &       & &  301.56 &  301.98 &  304.13 & &  299.98 &  298.69 &  299.24 &  297.81 &  298.7  \\        %
CF$_4$                                                                                                &  211.2 & 209.2 & &  212.35 &  212.94 &  216.14 & &  210.51 &  208.67 &  209.51 &  205.01 &  207.0  \\        %
PF$_3$                                                                                       &  176.8 & 177.7 & &  178.20 &  178.90 &  183.41 & &  173.76 &  173.08 &  174.25 &  172.08 &         \\        %
CFCl$_3$                                                                                              &        &       & &  120.02 &  120.79 &  124.94 & &  117.49 &  114.22 &  115.35 &  114.92 &  113.2  \\        %
NF$_3$                                                                                                &  -62.5 & -68.2 & &  -59.52 &  -58.33 &  -52.04 & &  -63.18 &  -67.70 &  -66.01 &  -74.86 &  -73.5  \\        %
F$_2$                                                                                                 & -293.4 &-299.8 & & -288.21 & -286.56 & -276.99 & & -292.54 & -298.33 & -295.93 & -307.14 & -296.3  \\[1.2ex] %
MAD                                                                                                   &    0.0 &       & &    2.4  &    2.8  &    6.6  & &  2.5    &         &         &         &         \\
MAD                                                                                                   &    2.4 &       & &    0.00 &    0.45 &    4.15 & &  1.63   &         &         &         &         \\
MAD                                                                                                   &        &   0.0 & &         &         &         & &         &    2.1  &    5.4  &    9.3  &         \\
\hline
\hline
\end{tabular}
\begin{tablenotes}
\item[a] See supplementary material for molecular geometries.
\item[b] QE shieldings from Ref.~\onlinecite{vasconcelos:jcp:2013}. %
\item[c] Molecular geometries and CASTEP shieldings from Ref.~\onlinecite{sadoc:pccp:2011} (except for PF$_3$).
\end{tablenotes}
\end{threeparttable}
\end{table*}

\begin{figure}
\includegraphics[width=0.48\textwidth]{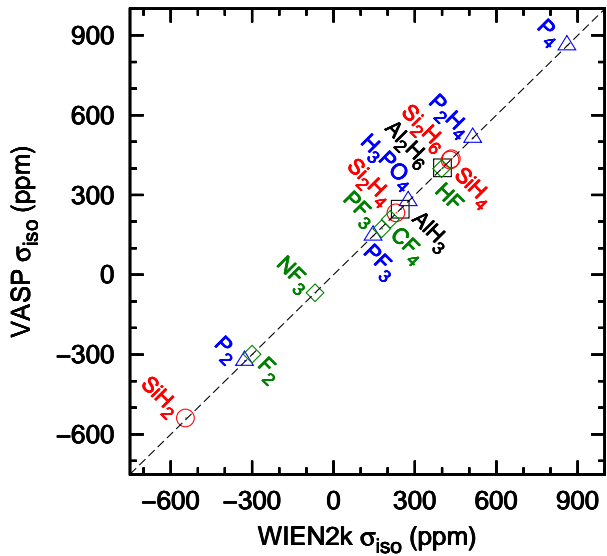}
\caption{\label{fig:molecular_shieldings_VASP_vs_WIEN2k}
(Color online) VASP versus WIEN2k PBE Al (black squares), Si (red cicles), P (blue triangles), and F (green diamonds) NMR shieldings.
Both VASP (with ``optimal'' PAW data sets) and WIEN2k results are scalar relativistic with $A$ and $B$ components, i.e.\ $^+B$.}
\end{figure}

Table~\ref{tab:molecular_shifts_dalton} compares chemical shieldings
from VASP calculations (with ``standard'' and with ``optimal''
PAW potentials) with all-electron WIEN2k results and DALTON all-electron
quantum chemical calculations. 
Here the VASP calculations were carried out in large boxes
($16 \times 16 \times 16$ to $17 \times 17 \times 17$~\AA$^3$),
in order to remove artificial fields from currents induced in the periodic
images and thus allow for comparison with the vacuum results from DALTON.
The WIEN2k results used boxes of $16 \times 16 \times 16$~\AA$^3$ only.

To avoid possible confusion with respect to differences in implementation of relativistic effects,
we excluded these in the comparison to DALTON results, {\it i.e.}, all DALTON results are non-relativistic,
and we have re-done the VASP calculations with PAW data sets generated from non-relativistic atomic calculations.
The non-relativistic results are labelled with NR in Table~\ref{tab:molecular_shifts_dalton}.

Table~\ref{tab:molecular_shifts_dalton} lists scalar-relativistic VASP and WIEN2k calculations of
the chemical shielding, as well.
As mentioned at the end of Sec.~\ref{subectGIPAW}, VASP calculations normally do not explicitly
take contributions from the ``small'' component into account (the ninth and tenth columns in
Table~\ref{tab:molecular_shifts_dalton}), whereas WIEN2k does.
To elucidate the effect of this approximation we have reconstructed the $B$-component of the 
scalar-relativistic atomic orbitals in VASP as well, and included their contribution to the chemical
shieldings in the column labelled $^+B$.

In Fig.~\ref{fig:molecular_shieldings_VASP_vs_WIEN2k} we compare WIEN2k and
VASP scalar-relativistic shieldings ($^+B$) for all compounds.
Overall the correlation is nearly perfect.
Below we discuss the differences in detail with the help of Table~\ref{tab:molecular_shifts_dalton}.

We start with the seemingly unambiguous cases, Al and Si. 
Agreement for non-relativistic results (columns ``NR'') between VASP
(optimal PAW data sets), WIEN2k, and best (aug-ccPCV5Z) DALTON 
shieldings is in general excellent.
Specifically for Al, the maximum difference between non-relativistic ``optimal''
VASP and the best DALTON shieldings is just 0.8~ppm (and the WIEN2k
values are very close, as well).

For Si, the maximum deviation between non-relativistic ``optimal'' VASP
and the best DALTON results increases to 4~ppm for the strongly deshielded
limit (SiH$_2$), which is still quite good.
We observe a progressive increase of the difference between VASP and DALTON
results with decreasing shielding.
The non-relativistic WIEN2k shieldings are 3 to 7~ppm lower than the results obtained with VASP,
and 2 to 3~ppm lower than the best DALTON results.

In the scalar-relativistic case, inclusion of the small component ($^+B$) has a
noticeable effect (compare the column nine [VASP, optim, $^+B$] and column ten [VASP, optim] in
Table~\ref{tab:molecular_shifts_dalton}):
in the high shielding limit, the shieldings differ by just 2-3~ppm,
whereas in the low shielding limit of inclusion of the $B$-component
reduces the shielding by 10~ppm (SiH$_2$).
The relativistic corrections to the shielding, calculated as the difference
between the two-component scalar-relativistic and non-relativistic shieldings,
are quite similar for VASP and WIEN2k: they differ by $\sim$~1~ppm.
This supports the validity of our two-component implementation.

Phosphorous is a more critical case.
Creating accurate PAW data sets for phosphorous, that include $2s$ and $2p$ states
as valence orbitals, is challenging.
The ``optimal'' P data set is created with $p$ core radii of 1.95~a.u.,
where the $2p$ core orbitals have almost negligible amplitudes (see Table~\ref{tab:pawdata}).
Nevertheless the results are excellent, with maximal differences between VASP and DALTON of 1~ppm.
Even for the difficult case of P$_2$, which is strongly deshielded, the agreement is very good.
The WIEN2k shieldings are 2-4~ppm lower than the corresponding VASP and DALTON results. 

Relativistic effects are increasingly more important towards the deshielded limit.
For P$_4$ relativistic effects amount to only a 0-1~ppm increase of the shielding,
whereas for P$_2$ the reduction is 22~ppm (with VASP as well as with WIEN2k).
Here again, inclusion of the small component ($^+B$) has a significant effect.
Neglecting explicit contributions from the small component of the scalar-relativistic
orbitals yields an increase of the P$_2$ shieldings of 10~ppm.
Including the contributions of the small component, the VASP and WIEN2k relativistic
corrections (difference between ``NR'' and ``$^+B$'') are identical to within $\sim 1$~ppm.

The agreement between VASP (``optim.'') and QE shieldings is good for high shieldings,
but becomes worse towards the low shielding limit.
We attribute this to the fact that in the QE calculations the $2s$ and $2p$ electrons
were treated as part of the core.
However, using the ``standard'' PAW data sets the VASP shieldings still significantly differ
from the QE results.
This illustrates to what extent the results depend on the particulars (matching radii, etc.)
of the ``standard'' PAW data sets.

The last molecular test systems are fluorine compounds.
Again agreement between VASP and DALTON is very good, but, as for the
silicon compounds, discrepancies with increasing deshielding are observed.
In general, VASP calculations tend to predict smaller shieldings
(and stronger deshielding), with differences of up to
4~ppm for PF$_3$, CFCl$_3$, NF$_3$ and F$_2$.
Here, carefully scrutinizing the convergence of the DALTON calculations
for F$_2$ and NF$_3$ shows that the basis set convergence using
Gaussian basis sets is slow, and differences between QZ and 5Z 
can be as large as 1.5~ppm (F$_2$).
In view of this it is not unlikely that the DALTON results might still be
inaccurate for these strongly deshielded cases (F$_2$ and NF$_3$),
despite the use of uncontracted 5Z basis sets. 
It has been suggested that relaxation of the F $1s$ core states might
play a role (see Ref.~\onlinecite{sadoc:pccp:2011}), {\it i.e.}, the frozen core approximation
in the GIPAW calculations may also explain part of the difference between the VASP and
DALTON results:
the close agreement between VASP (frozen core) and WIEN2k (all-electron)
results for F$_2$ and NF$_3$, however, does not support this.
In fact, calculations with WIEN2k show the F $1s$ contribution to be constant
within the series (305.8/306.4~ppm for non-/scalar-relativistic calculations)
In these calculations the F 1$s$ state self-consistently adapts to the
spherical part of the potential.
Admixture of $p$ and higher angular momentum states into the F 1$s$ is not possible,
but this should not really matter for a localized 1s core state at $-48$~Ry.
The WIEN2k and VASP fluorine shieldings agree very well, with (NR) differences less than 1~ppm,
except for
PF$_3$ where the (NR) difference is a bit larger (3~ppm).
Finally, agreement with the CASTEP GIPAW results of Sadoc {\it et al.}\cite{sadoc:pccp:2011}
is quite good as well
(our DALTON molecular quantum chemical shieldings are also very close to their
Gaussian basis set results).

All in all, Table~\ref{tab:molecular_shifts_dalton} shows that the ``optimal''
VASP potentials constitute a very stringent reference for future tests.
The agreement between VASP shieldings and the Gaussian based DALTON
calculations is excellent.
The shieldings obtained with WIEN2k are generally a few ppm smaller than those
obtained with VASP. This  might relate to the numerically virtually exact basis sets
close to the nucleus in the WIEN2k calculations.
For strongly deshielded cases (SiH$_2$, P$_2$, F$_2$) the
deviations between DALTON, WIEN2k, and VASP calculations are larger (up to 7 ppm). 
For these three cases the WIEN2k results are always the most negative ones.

For heavier elements, the treatment of the so-called semi-core states, e.g., the
low-lying $2p$ states of Si or P becomes quite challenging.
On the one hand, because of the very short bond-distances and anisotropic bonding situation in
these molecules the splitting of the $2p_x$, $2p_y$, and $2p_z$ states may
reach 10-25~mRy.
On the other hand the spin-orbit splitting of these states is
already more then twice as large. In the calculations presented above,
spin-orbit coupling has been neglected.
With WIEN2k it is possible to estimate the influence of SOC in the 2$p$-manifold
on the NMR shieldings: it turns out to be a fairly small effect.

Scalar relativistic effects can be substantial as can be the contribution
of the small-component, even for the light elements considered here.
However, often they appear to increase (in size) with reduced shielding.
So for calculating differences, which in practice is most important, we
think they can often be safely neglected.

In general, we observe that it is more difficult to predict and converge shieldings in the deshielded limit.
This is not unusual: in quantum chemical calculations, the diamagnetic contribution,
which depends only on the charge density, converges rapidly.
The paramagnetic contribution, which involves the Green's function and a sum over empty states, converges slower.
For fluorine, the diamagnetic contribution is approximately 500~ppm (DALTON number),
and only weakly dependent on the molecular composition.
The paramagnetic contribution goes from approximately $-$50~ppm for CH$_3$F
in the shielded limit to approximately $-$800~ppm
for F$_2$ in the deshielded limit. 
The latter exhibits a substantial variation with basis set size, and the variation of these contributions 
with the basis set size is an order of magnitude larger than the variation of the diamagnetic contribution.

We finish with a final look at the difference between standard and optimal
VASP PAW potentials.
For fluorine the agreement is excellent (except for the difficult strongly
deshielded cases).
This is not unexpected, since the same number of electrons are treated as valence. 
For Al(Si), the ``optimal'' PAW data sets give $\sim 7$(4)~ppm smaller shieldings
than the standard PAW potentials.
For P the optimal data sets give smaller shieldings in the high shielding
limit, and higher shieldings in the deshielded limit.
Since small systematic offsets matter little for comparison with experiment,
standard potentials will often give the right trends.
Indeed, for [AlH$_4$]$^-$ small deformations of the poly-anion result in very similar
changes in shielding for all PAW data sets and all quantum chemical basis sets.
However, in critical cases we advise to check against the optimal potentials,
because occasionally the trends can be broken.
This is illustrated by H$_3$PO$_4$, where the standard data set gives an
overestimation of the shielding by $\sim 10$~ppm.

\subsection{\label{sec:results:solids}Solids}

\begin{table*}
\caption{\label{tab:solid_shifts}Calculated isotropic F and O NMR shieldings $\sigma_{\rm iso}$ (in ppm),
calculated magnetic susceptibilities $\chi_{\rm m}$ (in $10^{-6}$\,cm$^3$\,mol$^{-1}$) and experimental chemical
shifts $\delta_{\rm iso}$ (in ppm).
GIPAW F and O shieldings calculated with other codes (column ``other GIPAW'') were taken
from Ref.~\onlinecite{sadoc:pccp:2011} and Refs.~\onlinecite{middlemiss:jmr:2010}
and \onlinecite{profeta:jacs:2004}, respectively.
Susceptibilities are per mole f.u.
Experimental susceptibilities are from Refs.~\onlinecite{L1,L2,L3,frederikse:pr:1966}.
Structures are from Ref.~\onlinecite{sadoc:pccp:2011} and the inorganic crystal structure database.\cite{laskowski:prb:2013,icsd}
All calculations are scalar relativistic with, inside the atomic spheres, one (VASP) and two (WIEN2k) component KS orbitals.
VASP numbers calculated with 900~eV (700~eV) kinetic energy cutoff in column optim.\ (standard).
``MAD'' denotes ``mean absolute deviation''.
}
\begin{tabular}{lddddddddd}
\hline
\hline
                           & \multicolumn{4}{c}{$\sigma_{\rm iso}$} & \multicolumn{1}{c}{$\delta_{\rm iso}$} & \multicolumn{4}{c}{$\chi_{\rm m}$} \\
\cline{2-5} \cline{7-10}
                           & \multicolumn{1}{c}{\multirow{2}{*}{WIEN2k}} & \multicolumn{2}{c}{VASP} & \multicolumn{1}{c}{other} & \multicolumn{1}{c}{\multirow{2}{*}{expt.}} & \multicolumn{1}{c}{\multirow{2}{*}{WIEN2k}} & \multicolumn{2}{c}{VASP} & \multicolumn{1}{c}{\multirow{2}{*}{expt.}} \\
\cline{3-4} \cline{8-9}
                           &         & \multicolumn{1}{c}{optim.} & \multicolumn{1}{c}{standard} & \multicolumn{1}{c}{GIPAW} & & & \multicolumn{1}{c}{optim.} & \multicolumn{1}{c}{standard} & \\
\hline                                 
\multicolumn{9}{l}{F shieldings} \\[1.2ex]                                           
NaF                        &  393.98 &  389.21 &  392.86 & 395.8\cite{sadoc:pccp:2011}      & -224.2\cite{sadoc:pccp:2011}   &  -16.0 &  -10.9 &  -11.6 & -15.6 \\ %
LiF                        &  370.12 &  368.14 &  367.65 & 369.3\cite{sadoc:pccp:2011}      & -204.3\cite{sadoc:pccp:2011}   &  -10.8 &   -9.1 &   -7.7 & -10.1 \\ %
InF$_3$                    &  365.55 &  363.17 &  365.02 &                                  & -209.2\cite{bureau:cp:1999}    &  -54.8 &  -43.4 &  -35.7 &       \\ %
MgF$_2$                    &  362.93 &  362.59 &  362.92 & 362.7\cite{sadoc:pccp:2011}      & -197.3\cite{sadoc:pccp:2011}   &  -23.5 &  -24.1 &  -38.8 & -22.7 \\ %
$\alpha$-AlF$_3$           &  335.32 &  334.06 &  334.51 &                                  & -172.0\cite{chupas:jacs:2001}  &  -30.1 &  -28.3 &  -24.9 & -13.9 \\ %
GaF$_3$                    &  312.51 &  307.23 &  310.63 &                                  & -171.2\cite{bureau:cp:1999}    &  -42.6 &  -29.5 &  -28.2 &       \\ %
KF                         &  271.08 &  270.41 &  271.27 & 268.1\cite{sadoc:pccp:2011}      & -133.3\cite{sadoc:pccp:2011}   &  -23.4 &  -25.1 &  -23.2 & -23.6 \\ %
RbF                        &  223.34 &  223.07 &  226.54 & 221.3\cite{sadoc:pccp:2011}      &  -90.9\cite{sadoc:pccp:2011}   &  -31.6 &  -31.0 &  -34.0 & -31.9 \\ %
CaF$_2$                    &  220.72 &  219.99 &  220.02 & 220.0\cite{sadoc:pccp:2011}      & -108.0\cite{sadoc:pccp:2011}   &  -25.8 &  -26.7 &  -22.7 & -28   \\ %
SrF$_2$                    &  216.17 &  216.05 &  220.17 & 215.3\cite{sadoc:pccp:2011}      &  -87.5\cite{sadoc:pccp:2011}   &  -34.4 &  -33.6 &  -34.6 & -37.2 \\ %
TlF                        &  148.92 &  146.10 &  152.06 &                                  &  -19.1\cite{gabuda:cpl:253}    &  -50.7 &  -30.6 &  -42.8 & -44.4 \\ %
CsF                        &  127.01 &  126.94 &  136.24 & 136.3\cite{sadoc:pccp:2011}      &  -11.2\cite{sadoc:pccp:2011}   &  -44.3 &  -40.9 &  -44.9 & -44.5 \\ %
BaF$_2$                    &  126.05 &  128.19 &  156.10 & 151.9\cite{sadoc:pccp:2011}      &  -14.3\cite{sadoc:pccp:2011}   &  -44.8 &  -42.2 &  -51.7 & -51   \\[1.2ex] %
MAD(F)                     &    0.00 &    1.76 &    4.41 &   5.0                            &                                &    0.0 &   4.9  &    6.4 &       \\
\hline
\multicolumn{9}{l}{O shieldings} \\[1.2ex]                                                                                                                            
BeO                        &  234.17 &  232.59 &  231.44 &  232.2\cite{middlemiss:jmr:2010} &  26.\cite{turner:jmr:1985}     & -12.6  & -11.2  & -10.8  & -11.9 \\ %
SiO$_2$                    &  214.21 &  213.83 &  213.95 &                                  &  41.\cite{profeta:jacs:2003}   & -24.3  & -24.9  & -23.7  & -28.6 \\ %
MgO                        &  201.77 &  200.25 &  200.82 &  198.0\cite{middlemiss:jmr:2010} &  47.\cite{turner:jmr:1985}     & -15.8  & -18.3  & -15.8  & -10.2 \\ %
BaSnO$_3$                  &   86.08 &   85.09 &   96.61 &   98.0\cite{middlemiss:jmr:2010} & 143.\cite{middlemiss:jmr:2010} & -73.1  & -61.5  & -70.4  &       \\ %
CaO                        & -145.56 & -146.05 & -145.30 & -156.6\cite{profeta:jacs:2004}   & 294.\cite{profeta:jacs:2004}   & -11.4  & -13.4  & -15.7  & -15.0 \\ %
BaZrO$_3$                  & -174.74 & -171.75 & -160.04 & -172.8\cite{middlemiss:jmr:2010} & 376.\cite{bastow:jpc:1996}     & -39.3  & -39.6  & -62.9  &       \\ %
SrO                        & -213.16 & -218.29 & -215.53 & -205.2\cite{middlemiss:jmr:2010} & 390.\cite{turner:jmr:1985}     & -16.5  & -17.6  & -22.4  & -35   \\ %
SrTiO$_3$                  & -290.61 & -289.75 & -289.14 & -287.3\cite{middlemiss:jmr:2010} & 465.\cite{bastow:jpc:1996}     & -10.0  &  -9.8  & -36.5  & -18.6 \\ %
\multirow{2}{*}{BaTiO$_3$} & -361.06 & -359.49 & -348.40 & -347.4\cite{middlemiss:jmr:2010} & 523.\cite{middlemiss:jmr:2010} & -12.4  & -11.0  & -48.4  &       \\ %
                           & -366.80 & -365.20 & -353.43 & -357.9\cite{middlemiss:jmr:2010} & 564.\cite{middlemiss:jmr:2010} &        &        &        &       \\ %
BaO                        & -481.43 & -483.71 & -458.46 & -444.3\cite{middlemiss:jmr:2010} & 629.\cite{turner:jmr:1985}     & -17.3  & -18.4  & -30.7  & -29.1 \\[1.2ex] %
MAD(O)                     &    0.00 &    1.76 &   7.48  &   10.2                           &                                &   0.0  &   2.2  &  11.5  &       \\
\hline
\hline
\end{tabular}
\end{table*}

\begin{figure}
\includegraphics[width=0.48\textwidth]{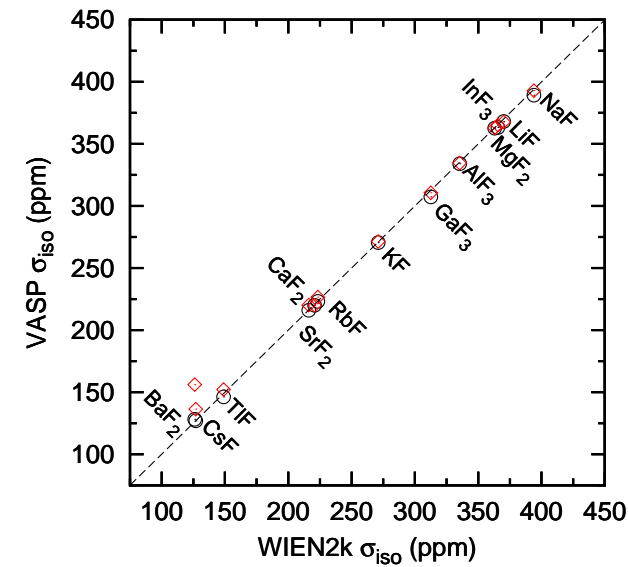}
\includegraphics[width=0.48\textwidth]{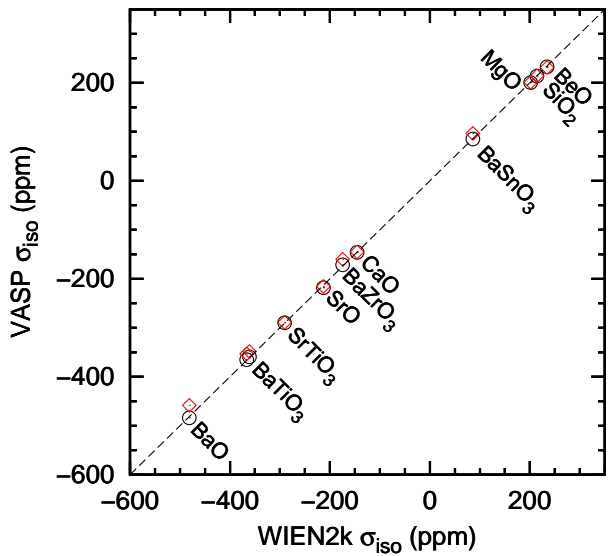}
\caption{\label{fig:solid_shieldings_PBE_VASP_vs_WIEN}
Comparison of shieldings calculated with WIEN2k and VASP with optimal (black circles) and standard (red diamonds) PAW data sets.
}
\end{figure}

\begin{figure}
\includegraphics[width=0.48\textwidth]{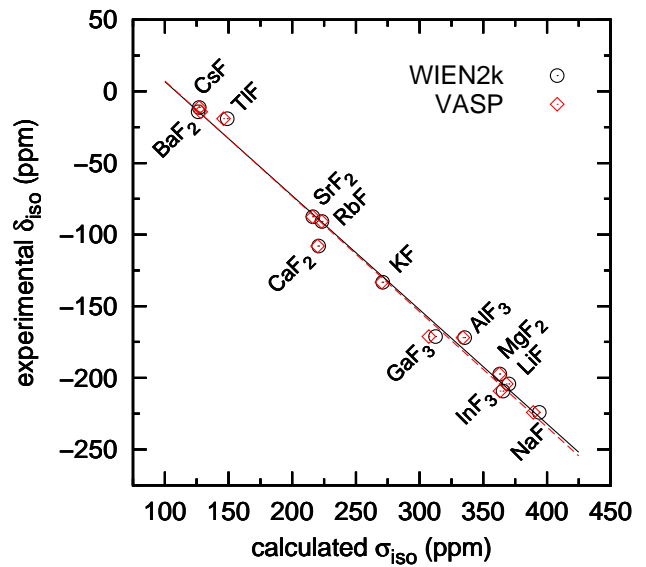}
\includegraphics[width=0.48\textwidth]{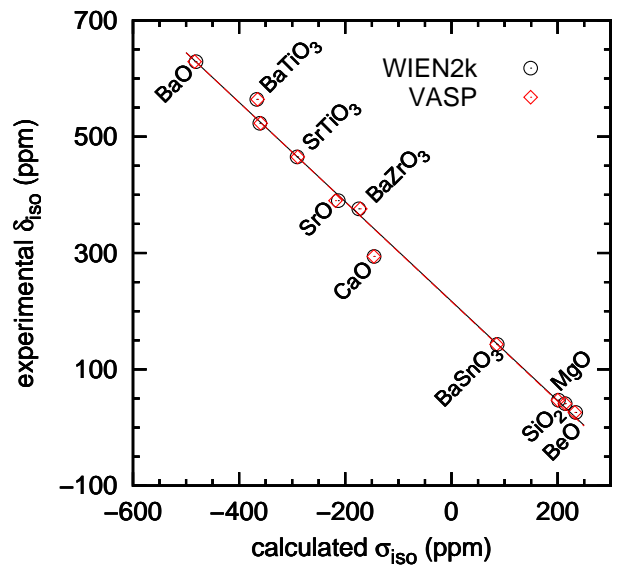}
\caption{\label{fig:solid_shieldings_PBE_VASP_WIEN_vs_exp}
Experimental shifts compared to calculated shieldings, for WIEN2k (black circles)
and VASP (red diamonds). The lines are linear fits, with parameters according
to Table~\ref{tab:regression}:black solid line corresponds to WIEN2k, dashed red line corresponds to VASP.
}
\end{figure}

Table~\ref{tab:solid_shifts} lists the isotropic F and O NMR shieldings and
magnetic susceptibilities for a range of fluoride and oxide systems,
calculated using VASP and WIEN2k.
As in the case of the molecular shieldings, using the high quality ``optimal'' PAW data sets,
the VASP and WIEN2k results are in very good agreement.
The largest deviations occur for KF, GaF$_3$ and SrO.
For GaF$_3$ this can be traced back to a difference in calculated susceptibilities.
Deviations are acceptable though, especially considering the shielding range of F and O.
This is illustrated in Fig.~\ref{fig:solid_shieldings_PBE_VASP_vs_WIEN}, by means of a plot of
VASP shieldings versus shieldings obtained with WIEN2k. 

Figure~\ref{fig:solid_shieldings_PBE_VASP_vs_WIEN} furthermore shows that on the scale of
the F and O shielding range in these compounds, using ``standard'' PAW data sets does not affect
the results appreciably, except for the Ba and Cs compounds.
For some of the other compounds there appear even to be slight improvements (e.g.\ for NaF).
All Cs and Ba data sets have unfrozen $5s$ and $5p$ semi-core states, i.e.\
the shallow core states are allowed to be polarized in the crystal field.
This, apparently, is not sufficient to get accurate shieldings.
The standard Ba data set has PAW matching radii of 2.8 (2.7) Bohr for the $s$ ($p$) channel,
with 2 projectors per channel.
The ``optimal'' set has  much smaller radii of 1.3-1.4 (2.2), and norm-conserving
pseudo partial waves for the $p$~channel.
Evidently this substantially reduces inaccuracies due to incompleteness in the PAW sphere,
which results in more accurate current densities that are ``felt'' in the induced field
at the neighboring nuclei.
With the ``optimal'' data sets, agreement with WIEN2k is very good.

In general the magnetic susceptibilities obtained with VASP using the ``optimal'' PAW data sets are in
fair agreement with the susceptibilities obtained with WIEN2k, with some exceptions for
compounds with heavier nuclei (TlF being the most dramatic).
The susceptibilities calculated using the ``standard'' PAW data sets in many cases are in
quite poor agreement with the all-electron WIEN2k results (e.g.\ SrTiO$_3$ and BaZrO$_3$).
This makes sense, as the YPM expression for the magnetic susceptibility lacks one-centre corrections
(Eqs.~47 and 48 of Ref.~\onlinecite{yates:prb:2007}), and is expected to become more accurate
when PAW data sets become harder (smaller core radii and/or norm-conserving).
Shieldings calculated discarding the susceptibility contribution of Eq.~\ref{eq.bg0} are listed as supplementary material.

In Fig.~\ref{fig:solid_shieldings_PBE_VASP_WIEN_vs_exp} we compare WIEN2k and VASP shieldings
(``optimal'' PAW data sets) to chemical {\it shifts} from experiment.
A fit is done, according to:
\begin{equation}
\delta_{\rm iso}^{\rm exp} = \sigma_{\rm ref} - m \sigma_{\rm iso}^{\rm calc} ~.
\label{eq:reference}
\end{equation}
Fit results are in Table~\ref{tab:regression}. 
Note that the Ca compounds are well besides the fitted straight lines.
Indeed, it is known that the empty Ca $3d$-states are too close to the valence band
maximum in DFT, resulting in a deviation of the O shift.\cite{profeta:jacs:2004}

\begin{table}
\caption{Fit parameters according to Eq.~\protect \ref{eq:reference} with standard errors in brackets and Pearson correlation coefficient $r$.}
\begin{tabular}{lrrr}
\hline
\hline
       & \multicolumn{1}{c}{$\sigma_{\rm ref}$} & \multicolumn{1}{c}{$m$} & \multicolumn{1}{c}{$r$} \\
\hline
\multicolumn{3}{l}{fluorides} \\[1.2ex]
WIEN2k           &  86.47(7.08) & $-$0.7964(0.0250) & $-$0.9946 \\          %
VASP (optim)     &  87.76(7.22) & $-$0.8056(0.0257) & $-$0.9945 \\          %
VASP (standard)  & 101.64(9.54) & $-$0.8429(0.0337) & $-$0.9914 \\[1.2ex]   %
\multicolumn{3}{l}{oxides} \\[1.2ex]
WIEN2k           & 217.23(6.77) & $-$0.8546(0.0247) & $-$0.9963 \\          %
VASP (optim)     & 216.67(7.00) & $-$0.8558(0.0256) & $-$0.9960 \\          %
VASP (standard)  & 220.65(7.79) & $-$0.8724(0.0292) & $-$0.9950 \\          %
\hline
\hline
\end{tabular}
\label{tab:regression}
\end{table}

The fit of Eq.~\ref{eq:reference} relates the calculated shieldings to chemical shifts from experiment.
Table IV shows that the slope $m$ amd reference shielding $\sigma_{\rm ref}$
from VASP calculations with optimal PAW data sets and as obtained with WIEN2k are in excellent agreement.

\section{\label{sec:conclusions}Conclusions}

In the calculation of NMR shieldings for the molecular systems considered in this paper
we have pushed the convergence of the results of the DALTON (Gaussian basis set), WIEN2k (APW+lo),
and VASP (GIPAW) calculations with respect to their basis sets as far as practicably possible.
If this is done, generally excellent agreement can be obtained for very different codes and implementations. 
Specifically, for Si, Al and P agreement of GIPAW calculations (VASP) with the all-electron Gaussian basis set
results (DALTON) is excellent, with inaccuracies of several ppm in the extreme deshielded limit for SiH$_2$.
For F, in the deshielded range, differences of up to 4 ppm occur,
but in this case the DALTON calculations are probably still not fully converged with respect to the basis set size,
and the VASP results are validated by the comparison to WIEN2k.
Generally, the all-electron APW+lo WIEN2k shieldings agree very well with those obtained from DALTON and VASP calculations,
although the WIEN2k shieldings are consistently slightly lower than the DALTON (and most of the VASP) results.
This might be due to the superior quality of the APW+lo basis sets of WIEN2k close to the nuclei.

Scalar relativistic effects, even for light nuclei, can be substantial.
They increase towards the deshielded limit, where inclusion of the
small-component of the wave function can considerably affect the shielding.

In general, the agreement between VASP and WIEN2k shieldings is very good for the molecular and the 
solid state systems considered here.
We consider this to be a validation of our norm-conserving and hard GW PAW data sets.
Completeness of the projector functions/partial waves inside the PAW spheres is crucially important,
not only for the elements for which the shielding is calculated, but for other atomic constituents as well:
This is especially true for Cs and Ba, that have very shallow semi-core states.
Large matching radii in the pseudization of Cs and Ba (as regularly used) yield wrong oxygen and fluorine shifts.
Reducing these radii and the use of norm-conserving partial waves yields a more accurate description
of the current density already on the plane wave grid, {\it i.e.},
the results are less affected by undercompleteness of the PAW one-center basis.
This, however, generally comes at the price of an increase in the cutoff energy of the plane wave basis set.
Standard VASP PAW data sets, although being less accurate, describe trends and chemical differences quite well.
Hence, they can be used in shielding calculations, although for small differences double checks
with more accurate PAW data sets are in order.
In general, it is more difficult to predict and converge shieldings in the
deshielded limit, where the paramagnetic contribution, that involves the
Green's function, is larger.

Harder data sets give, in general, better values for the susceptibilities,
since one-centre corrections are presently missing for the susceptibilities.
In most cases reasonable susceptibilities can often be obtained with standard
data sets already, and the contribution
of the susceptibility to the shielding is, in general, modest anyway.
This is crucial, since our, as well as other implementations use only one-centre corrections for the current density.

\section*{supplementary material}
See supplementary material for the structure of the Al, Si and P containing molecules and solid state shieldings calculated without the $G=0$ contribution.

\begin{acknowledgments}
We thank Dr.~F.~M.~Vasconcelos for useful discussions.
The work was partly supported by the Austrian Science Fund (FWF) within the 
Spezialforschungsbereich Vienna Computational Materials Laboratory (SFB
ViCoM, F41).
The work of
GAW is part of the research programme of the
``Stichting voor Fundamenteel Onderzoek der Materie (FOM)'',
which is financially supported by the
``Nederlandse Organisatie voor Wetenschappelijk Onderzoek (NWO)''.
\end{acknowledgments}


\begin{thebibliography}{10}

\bibitem{helgaker:cr:1999}
{T.~Helgaker, M.~Jaszu\'{n}ski, and K.~Ruud},
\newblock Chem.~Rev.~{\bf 99}, 293 (1999).

\bibitem{mauri:prl:1996}
{F.~Mauri, B.~G.~Pfrommer, and S.~G.~Louie},
\newblock Phys.~Rev.~Lett.~{\bf 77}, 5300 (1996).

\bibitem{pickard:prb:2001}
{C.~J.~Pickard and F.~Mauri},
\newblock Phys.~Rev.~B~{\bf 63}, 245101 (2001).

\bibitem{yates:prb:2007}
{J.~R.~Yates, C.~J.~Pickard and F.~Mauri},
\newblock Phys.~Rev.~B~{\bf 76}, 024401 (2007).

\bibitem{bloechl:prb:1994}
{P.~E.~Bl\"{o}chl},
\newblock Phys.~Rev.~B~{\bf 50}, 17953 (1994).

\bibitem{ditchfield:mp:1974}
{R.~Ditchfield},
\newblock Mol.~Phys.~{\bf 27}, 789 (1974).

\bibitem{castep}
{S.~J.~Clark, M.~D.~Segall, C.~J.~Pickard, P.~J.~Hasnip, M.~J.~Probert,
K.~Refson and M.~C.~Payne},
\newblock Z.~Kristallogr. {\bf 220}, 567 (2005).

\bibitem{espresso}
{Paolo Giannozzi, Stefano Baroni, Nicola Bonini, Matteo Calandra, Roberto Car,
Carlo Cavazzoni, Davide Ceresoli, Guido L Chiarotti, Matteo Cococcioni,
Ismaila Dabo, Andrea {Dal Corso}, Stefano de Gironcoli, Stefano Fabris,
Guido Fratesi, Ralph Gebauer, Uwe Gerstmann, Christos Gougoussis,
Anton Kokalj, Michele Lazzeri, Layla Martin-Samos, Nicola Marzari,
Francesco Mauri, Riccardo Mazzarello, Stefano Paolini, Alfredo Pasquarello,
Lorenzo Paulatto, Carlo Sbraccia, Sandro Scandolo, Gabriele Sclauzero,
Ari P Seitsonen, Alexander Smogunov, Paolo Umari, and Renata M Wentzcovitch},
\newblock J.~Phys.:~Condens.~Matter {\bf 21}, 395502 (2009), http://www.quantum-espresso.org

\bibitem{charpentier:ssnmr:2011}
{T.~Charpentier},
\newblock Solid State Nucl.~Magn.~Reson.~{\bf 40}, 1 (2011).

\bibitem{laskowski:prb:2012}
{R.~Laskowski and P.~Blaha},
\newblock Phys.~Rev.~B {\bf 85}, 035132 (2012).

\bibitem{laskowski:prb:2013}
{R.~Laskowski, P.~Blaha, and F.~Tran},
\newblock Phys.~Rev.~B {\bf 87}, 195130 (2013).

\bibitem{laskowski:prb:2014}
{R.~Laskowski and P.~Blaha},
\newblock Phys.~Rev.~B {\bf 89}, 014402 (2014).

\bibitem{vasconcelos:jcp:2013}
{F.~Vasconcelos, G.~A.~de Wijs, R.~W.~A.~Havenith, M.~Marsman, and G.~Kresse},
\newblock J.~Chem.~Phys.~{\bf 139}, 014109 (2013).

\bibitem{dalton}
Dalton, a molecular electronic structure program,
Release Dalton2011 (2011), see http://daltonprogram.org

\bibitem{louie:prl:1996}
{F.~Mauri and S.~G.~Louie},
\newblock Phys.~Rev.~Lett.~{\bf 76}, 4246 (1996).

\bibitem{rappe:prb:1990}
{A.~M.~Rappe, K.~M.~Rabe, E.~Kaxiras, and J.~D.~Jaonnopoulos},
\newblock Phys.~Rev.~B~{\bf 41}, 1227 (1990).

\bibitem{kresse:jpcm:1994}
{G.~Kresse and J.~Hafner},
\newblock J.~Phys.:~Condens.~Matter~{\bf 6}, 8245 (1994).

\bibitem{kresse:prb:1999}
{G.~Kresse and D.~Joubert},
\newblock Phys.~Rev.~B~{\bf 59}, 1758 (1999).

\bibitem{gregor:jcp:1999}
{T.~Gregor, F.~Mauri, and R.~Car},
\newblock J.~Chem.~Phys.~{\bf 111}, 1815 (1999).

\bibitem{pbe1}
{J.~P.~Perdew, K.~Burke, and M.~Ernzerhof},
\newblock Phys.~Rev.~Lett. {\bf 77}, 3865 (1996).

\bibitem{pbe2}
{J.~P.~Perdew, K.~Burke, and M.~Ernzerhof},
\newblock Phys.~Rev.~Lett. {\bf 78}, 1396 (1997).

\bibitem{vasp:manual}
{G.~Kresse, M.~Marsman, and J.~Furthm{\"{u}}ller},
{\it VASP the Guide} (Vienna, March~26, 2015),
http://cms.mpi.univie.ac.at/vasp/vasp/vasp.html

\bibitem{klimes:prb:2014}
{J.~Klime{\v{s}}, M.~Kaltak, and G.~Kresse},
\newblock Phys.~Rev.~B {\bf 90}, 075125 (2014).

\bibitem{zora}
{E.~van~Lenthe, E.~J.~Baerends, and J.~G.~Snijders},
\newblock J.~Chem.~Phys. {\bf 99}, 4597 (1993).

\bibitem{daltondft}
{T.~Helgaker, P.~J.~Wilson, R.~D.~Amos, and N.~C.~Handy},
\newblock J.~Chem.~Phys. {\bf 113}, 2983 (2000).

\bibitem{dunning1}
{T.~H.~Dunning,~Jr.},
\newblock J.~Chem.~Phys. {\bf 90}, 1007 (1989).

\bibitem{dunning2}
{D~.E.~Woon, and T.~H.~Dunning,~Jr.},
\newblock (to be published).

\bibitem{dunning3}
{D~.E.~Woon, and T.~H.~Dunning,~Jr.},
\newblock J.~Chem.~Phys. {\bf 98}, 1358 (1993). 

\bibitem{dunning4}
{D~.E.~Woon, and T.~H.~Dunning,~Jr.},
\newblock J.~Chem.~Phys. {\bf 103}, 4572 (1995). 

\bibitem{dunning5}
{K.~A.~Peterson, and T.~H.~Dunning,~Jr.},
\newblock J.~Chem.~Phys. {\bf 117}, 10548 (2002).

\bibitem{dunning6}
{R.~A.~Kendall, T.~H.~Dunning,~Jr. and R.~J.~Harrison},
\newblock J.~Chem.~Phys. {\bf 96}, 6796 (1992).

\bibitem{dunning7}
{D.~Feller},
\newblock (unpublished).

\bibitem{dunning8}
{S.~Mielke},
\newblock (unpublished).

\bibitem{emsl1}
{D.~Feller},
\newblock J.~Comp.~Chem. {\bf 17}, 1571 (1996).

\bibitem{emsl2}
{K.~L.~Schuchardt, B.~T.~Didier, T.~Elsethagen, L.~Sun, V.~Gurumoorthi, J.~Chase, J.~Li, and T.~L.~Windus},
\newblock J.~Chem.~Inf.~Model. {\bf 47}, 1045 (2007).

\bibitem{sadoc:pccp:2011}
{A.~Sadoc, M.~Body, C.~Legein, M.~Biswal, F.~Fayon, X.~Rocquefelte, and F.~Boucher},
\newblock Phys.~Chem.~Chem.~Phys.~{\bf 13}, 18539 (2011).

\bibitem{middlemiss:jmr:2010}
{D.~S.~Middlemiss, F.~Blanc, C.~J.~Pickard, and C.~P.~Grey},
\newblock J.~Magn.~Reson.~{\bf 204}, 1 (2010).

\bibitem{profeta:jacs:2004}
{M.~Profeta, M.~Benoit, F.~Mauri, and C.~J.~Pickard},
\newblock J.~Am.~Chem.~Soc.~{\bf 126}, 12628 (2004).

\bibitem{L1}
{\sl Landolt-B{\"{o}}rnstein, Numerical Data and Functional Relationships in Science
and Technology, New Series, II/16, Diamagnetic Susceptibility}
\newblock (Springer, Heidelberg, 1986).

\bibitem{L2}
{\sl Landolt-B{\"{o}}rnstein, Numerical Data and Functional Relationships in Science
and Technology, New Series, II/2, II/8, II/10, II/11, II/12a,
Coordination and Organometallic Transition Metal Compounds}
\newblock (Springer, Heidelberg, 1986).

\bibitem{L3}
{\sl Tables de Constantes et Donn{\'{e}}es Num{\'{e}}rique, Volume 7, Relaxation Paramagn{\'{e}}tique}
\newblock (Masson, Paris, 1957).

\bibitem{frederikse:pr:1966}
{H.~P.~R.~Frederikse, and G.~A.~Candela},
\newblock Phys.~Rev. {\bf 147}, 583 (1966).

\bibitem{icsd}
http://icsd.fiz-karlsruhe.de/icsd

\bibitem{bureau:cp:1999}
{B.~Bureau, G.~Silly, J.~Buzar{\'{e}}, and J.~Emery},
\newblock Chem.~Phys. {\bf 249}, 89 (1999).

\bibitem{chupas:jacs:2001}
{P.~J.~Chupas, M.~F.~Ciraolo, J.~C.~Hanson, and C.~P.~Grey},
\newblock J.~Am.~Chem.~Soc. {\bf 123}, 1694 (2001).

\bibitem{gabuda:cpl:253}
{S.~Gabuda, S.~Kozlova, and R.~Davidovich},
\newblock Chem.~Phys.~Lett. {\bf 263}, 263 (1996).

\bibitem{turner:jmr:1985}
{G.~L.~Turner, S.~E.~Chung, and E.~Oldfield},
\newblock J.~Magn.~Reson.~{\bf 64}, 316 (1985).

\bibitem{profeta:jacs:2003}
{M.~Profeta, F.~Mauri, and C.~J.~Pickard},
\newblock J.~Am.~Chem.~Soc. {\bf 125}, 541 (2003).

\bibitem{bastow:jpc:1996}
{T.~J.~Bastow, P.~J.~Dirken, M.~E.~Smith, and H.~J.~Whitfield},
\newblock J.~Phys.~Chem.~{\bf 100}, 18539 (1996).

\end{thebibliography}
\end{document}

\title{%
NMR shieldings from density functional perturbation theory:\\
GIPAW versus all-electron calculations.\\[1.2ex]
Supplementary material
}

\author{G.~A.~de Wijs}
\affiliation{%
Radboud University, Institute for Molecules and Materials,
Heyendaalseweg 135, NL-6525 AJ Nijmegen, The Netherlands
}

\author{R.~Laskowski}
\affiliation{%
Institute of High Performance Computing, A$\ast$STAR, 1 Fusionopolis Way, \#16-16, Connexis, Singapore 138632
}

\author{P.~Blaha}
\affiliation{%
Institute of Materials Chemistry, Vienna University of Technology,
Getreidemarkt 9/165-TC, A-1060 Vienna, Austria
}

\author{R.~W.~A.~Havenith}
\affiliation{%
Zernike Institute for Advanced Materials, Stratingh Institute for Chemistry, University of Groningen,
Nijenborgh~4, NL-9747~AG Groningen, The Netherlands
}
\affiliation{%
Ghent Quantum Chemistry Group, Department of Inorganic and Physical Chemistry,
Ghent University, Krijgslaan 281 (S3), B-9000 Gent, Belgium
}

\author{G.~Kresse}
\author{M.~Marsman}
\affiliation{%
Faculty of Physics and Center for Computational Materials Science,
University of Vienna, Sensengasse 8/12, A-1090 Vienna, Austria
}

\pacs{}

\keywords{}
\maketitle

Calculated shieldings calculated without the susceptibility contribution are in
Table~\ref{tab:solid_shifts_nochi}.
Coordinates of the atoms of the molecules with Al, Si and P nuclei are in Tables~\ref{tab:Al}, \ref{tab:Si} and \ref{tab:P} respectively.

\begin{table}
\caption{\label{tab:solid_shifts_nochi}Isotropic NMR shieldings without $G=0$ contribution (ppm).
Other details as in Table~III.
}

\begin{tabular}{lddd}
\hline
\hline
                             & \multicolumn{1}{c}{\multirow{2}{*}{WIEN2k}} & \multicolumn{2}{c}{VASP} \\
\cline{3-4}
                             &         & \multicolumn{1}{c}{optim.} & \multicolumn{1}{c}{standard} \\
\hline

\multicolumn{4}{l}{F shieldings}                             \\[1.2ex]
 NaF                         &  385.04 &  383.14  &  386.38  \\
 LiF                         &  360.95 &  360.37  &  361.10  \\
 InF$_3$                     &  353.01 &  353.22  &  356.83  \\
 MgF$_2$                     &  352.78 &  352.16  &  354.54  \\
 $\alpha$-AlF$_3$            &  325.72 &  325.03  &  326.57  \\
 GaF$_3$                     &  299.85 &  298.47  &  302.27  \\
 KF                          &  262.64 &  261.37  &  262.91  \\
 RbF                         &  213.55 &  213.45  &  215.99  \\
 CaF$_2$                     &  211.92 &  210.88  &  212.26  \\
 SrF$_2$                     &  206.32 &  206.43  &  210.28  \\
 TlF                         &  132.65 &  136.26  &  138.32  \\
 CsF                         &  115.77 &  116.55  &  124.85  \\
 BaF$_2$                     &  115.56 &  118.34  &  144.00  \\[1.2ex]
MAD(F)                       &    0.00 &    1.16  &    4.66  \\
\hline
\multicolumn{4}{l}{O shieldings}                             \\[1.2ex]

 BeO                         &  221.48 &  221.24  &  220.56  \\
 SiO$_2$                     &  205.26 &  204.64  &  205.21  \\
 MgO                         &  190.04 &  186.67  &  189.09  \\
 BaSnO$_3$                   &   71.47 &   72.81  &   82.55  \\
 CaO                         & -151.26 & -152.79  & -153.15  \\
 BaZrO$_3$                   & -182.14 & -179.20  & -171.87  \\
 SrO                         & -220.06 & -225.67  & -224.90  \\
 SrTiO$_3$                   & -292.89 & -291.97  & -297.44  \\
 \multirow{2}{*}{BaTiO$_3$}  & -363.75 & -361.88  & -358.88  \\
                             & -369.49 & -367.59  & -363.91  \\
 BaO                         & -487.13 & -489.80  & -468.58  \\[1.2ex]
MAD(O)                       &    0.00 &    2.09  &    5.78  \\
\hline
\hline
\end{tabular}

\end{table}

\begin{table}

\caption{Atomic positions for molecules containing Al nuclei.}
\begin{tabular}{lrrrr}
\hline
\hline
               &     & \multicolumn{1}{c}{$x$ (\AA)} & \multicolumn{1}{c}{$y$ (\AA)} & \multicolumn{1}{c}{$z$ (\AA)} \\
\hline
[AlH$_4$]$^-$  & Al  &  3.499929 &  3.499929 &  3.147370\\
               &  H  &  4.434678 &  2.565180 &  2.189269\\
               &  H  &  2.565180 &  2.565180 &  4.105461\\
               &  H  &  2.565180 &  4.434678 &  2.189269\\
               &  H  &  4.434678 &  4.434678 &  4.105461\\[1.2ex]

[AlH$_4$]$^-$  & Al  &    3.538681 &    2.270439 &    5.289092\\
               &  H  &    2.224501 &    3.237801 &    5.289092\\
               &  H  &    4.852860 &    3.237801 &    5.289092\\
               &  H  &    3.538681 &    1.287091 &    3.994212\\
               &  H  &    3.538681 &    1.287091 &    6.583967\\[1.2ex]

[AlH$_4$]$^-$  & Al  &    3.367010 &    3.166802 &    2.160921\\
               &  H  &    3.367010 &    3.378023 &     .546772\\
               &  H  &    2.071760 &    3.912999 &    2.824731\\
               &  H  &    4.662261 &    3.912999 &    2.824731\\
               &  H  &    3.367010 &    1.583859 &    2.505351\\[1.2ex]

[AlH$_4$]$^-$  & Al  &    6.431564 &    2.961089 &    3.727782\\
               &  H  &    7.537068 &    2.325728 &    4.763868\\
               &  H  &    5.326061 &    3.596451 &    4.763868\\
               &  H  &    7.133835 &    4.109440 &    2.835542\\
               &  H  &    5.729294 &    1.812738 &    2.835542\\[1.2ex]

Al$_2$H$_6$    & Al  &    1.290911 &     .000000 &     .000000\\
               & Al  & $-$1.290911 &     .000000 &     .000000\\
               &  H  &     .000000 &     .000000 &    1.160480\\
               &  H  &     .000000 &     .000000 & $-$1.160480\\
               &  H  &    1.971661 &    1.430667 &     .000000\\
               &  H  &    1.971661 & $-$1.430667 &     .000000\\
               &  H  & $-$1.971661 &    1.430667 &     .000000\\
               &  H  & $-$1.971661 & $-$1.430667 &     .000000\\[1.2ex]

AlH$_3$        & Al  &     .000000 &     .000000 &     .000000\\
               &  H  &     .000000 &    1.579020 &     .000000\\
               &  H  &    1.367472 &  $-$.789511 &     .000000\\
               &  H  & $-$1.367472 &  $-$.789511 &     .000000\\
\hline
\hline
\end{tabular}
\label{tab:Al}

\end{table}

\begin{table}

\caption{Atomic positions for molecules containing Si nuclei.}
\begin{tabular}{lrrrr}
\hline
\hline
               &     & \multicolumn{1}{c}{$x$ (\AA)} & \multicolumn{1}{c}{$y$ (\AA)} & \multicolumn{1}{c}{$z$ (\AA)} \\
\hline
Si$_2$H$_6$   &  Si  &     .000000 &     .000000 &    1.174799\\
              &  Si  &     .000000 &     .000000 & $-$1.174799\\
              &   H  &     .000000 &    1.389598 &    1.691101\\
              &   H  & $-$1.203401 &  $-$.694799 &    1.691101\\
              &   H  &    1.203401 &  $-$.694799 &    1.691101\\
              &   H  &     .000000 & $-$1.389598 & $-$1.691101\\
              &   H  & $-$1.203401 &     .694799 & $-$1.691101\\
              &   H  &    1.203401 &     .694799 & $-$1.691101\\[1.2ex]
                                                      
SiH$_4$       &  Si  &     .000000 &     .000000 &     .000000\\
              &   H  &     .861738 &  $-$.861738 &     .861738\\
              &   H  &  $-$.861738 &     .861738 &     .861738\\
              &   H  &  $-$.861738 &  $-$.861738 &  $-$.861738\\
              &   H  &     .861738 &     .861738 &     .861738\\[1.2ex]
                                                      
Si$_2$H$_4$   &  Si  &    1.052200 &     .000000 &     .217920\\
              &  Si  & $-$1.052200 &     .000000 &  $-$.217920\\
              &   H  &    1.838784 &    1.244259 &  $-$.026953\\
              &   H  &    1.838784 & $-$1.244259 &  $-$.026953\\
              &   H  & $-$1.838784 &    1.244259 &     .026953\\
              &   H  & $-$1.838784 & $-$1.244259 &     .026953\\[1.2ex]
                                                      
SiH$_2$       &  Si  &     .000000 &     .000000 &     .132600\\
              &   H  &     .000000 &    1.091200 &  $-$.928097\\
              &   H  &     .000000 & $-$1.091200 &  $-$.928097\\
\hline
\hline
\end{tabular}
\label{tab:Si}

\end{table}

\begin{table}

\caption{Atomic positions for molecules containing P nuclei.}
\begin{tabular}{lrrrr}
\hline
\hline
               &     & \multicolumn{1}{c}{$x$ (\AA)} & \multicolumn{1}{c}{$y$ (\AA)} & \multicolumn{1}{c}{$z$ (\AA)} \\
\hline
P$_4$                     &     P &     .772319 &     .772319 &     .772319\\
                          &     P &  $-$.772319 &  $-$.772319 &     .772319\\
                          &     P &  $-$.772319 &     .772319 &  $-$.772319\\
                          &     P &     .772319 &  $-$.772319 &  $-$.772319\\[1.2ex]
                                                                   
PH$_3$                    &     P &     .000000 &     .000000 &     .000000\\
                          &     H &     .945009 &     .508249 &     .945009\\
                          &     H &     .945009 &  $-$.945009 &  $-$.508249\\
                          &     H &  $-$.508249 &  $-$.945009 &     .945009\\[1.2ex]
                                                                   
P$_2$H$_4$                &     P &     .000000 &    1.119312 &  $-$.085740\\
                          &     P &     .000000 & $-$1.119312 &  $-$.085740\\
                          &     H &  $-$.191443 &    1.364852 &    1.296710\\
                          &     H &    1.409764 &    1.237215 &  $-$.010603\\
                          &     H &     .191443 & $-$1.364852 &    1.296710\\
                          &     H & $-$1.409764 & $-$1.237215 &  $-$.010603\\[1.2ex]
                                                                   
H$_3$PO$_4$               &     P &     .000000 &     .000000 &     .000000\\
                          &     O &  $-$.706579 &    1.156170 &  $-$.569987\\
                          &     O &  $-$.893018 &  $-$.897538 &    1.007560\\
                          &     O &     .581529 & $-$1.121920 &  $-$.991408\\
                          &     O &    1.325710 &     .392790 &     .804032\\
                          &     H &  $-$.069470 & $-$1.407680 & $-$1.657030\\
                          &     H & $-$1.593160 &  $-$.358900 &    1.418850\\
                          &     H &    1.824050 &  $-$.371290 &    1.145860\\[1.2ex]
                                                                   
PF$_3$                    &     P &     .000000 &     .000000 &     .000000\\
                          &     F &     .000000 &    1.370150 &  $-$.790302\\
                          &     F &    1.187690 &  $-$.686336 &  $-$.787314\\
                          &     F & $-$1.187690 &  $-$.686336 &  $-$.787314\\[1.2ex]
                                                                   
P$_2$                     &     P &     .000000 &     .000000 &     .000000\\
                          &     P &     .000000 &     .000000 &    1.890458\\[1.2ex]
\hline
\hline
\end{tabular}
\label{tab:P}

\end{table}